\documentclass[sts]{imsart}

\RequirePackage[colorlinks,citecolor=blue,urlcolor=blue]{hyperref}
\usepackage{amsthm,amsmath,amssymb,amsfonts,natbib,bm,cases,color}
\usepackage[pdftex]{graphicx}
\usepackage[margin=3.0cm]{geometry}
\RequirePackage[colorlinks,citecolor=blue,urlcolor=blue]{hyperref}

\startlocaldefs

\numberwithin{equation}{section}
\theoremstyle{plain}

\newtheorem{proposition}{Proposition}[section]
\theoremstyle{definition}
\newtheorem{definition}{Definition}
\newtheorem{example}[definition]{Example}

\endlocaldefs

\begin{document}

\begin{frontmatter}
\title{Statistical Inference with Different Missing-data Mechanisms}
\runtitle{Different Missing-data Mechanisms}

\begin{aug}
\author{\fnms{Kosuke} \snm{Morikawa}\thanksref{t1}\ead[label=e1]{morikawa@sigmath.es.osaka-u.ac.jp}
},
\author{\fnms{Yutaka} \snm{Kano}\thanksref{t1}\ead[label=e2]{kano@sigmath.es.osaka-u.ac.jp}
}
\runauthor{Morikawa and Kano}
\address{Division of Mathematical Science, Graduate School of Engineering Science, \\
Osaka University, Toyonaka, Osaka 560-8571, Japan.}
\affiliation{Osaka University}
\thankstext{t1}{\printead{e1,e2}}
\end{aug}
\begin{abstract}
When data are missing due to at most one cause from some time to next time, we can make sampling distribution inferences about the parameter of the data by modeling the missing-data mechanism correctly. Proverbially, in case its mechanism is missing at random (MAR), it can be ignored, but in case not missing at random (NMAR), it can not be. There are no methods, however, to analyze when missing of the data can occur because of several causes despite of there being many such data in practice. Hence the aim of this paper is to propose how to inference on such data. Concretely, we extend the missing-data indicator from usual binary random vectors to discrete random vectors, define missing-data mechanism for every causes and research ignorability of a mixture of missing-data mechanisms such as ``MAR \& MAR" and ``MAR \& NMAR". In particular, when the combination of mechanisms is ``MAR \& NMAR", generally the component of MAR can not be ignored, but in special case, it can be.
\end{abstract}
\end{frontmatter}
\section{Introduction}
There are a lot of theory to analyze data which are missing by only one cause.  When analyzing such data, it is very important to specify the missing-data mechanism: when  a missing-data mechanism is missing completely at ransom(MCAR) or missing at random(MAR), we do not have to specify it correctly, however, when it is not missing at random(NMAR), we have to specify it correctly(\citet{little02}). In some cases, it is conceivable that data would be missing by several different causes. Recently, when data are missing, their missing causes are usually disclosed by follow-up survey and this type of data are reported as paradata, which are data about the process by which they are collected. For example, in clinical trials, \citet{machin88} reported results of comparative trial of two dosages of depot medroxyprogesterone acetate (DMPA, 100mg and 150mg) in which subjects were missing over 40\% at the endpoint. In this data, all missing causes are revealed and listed such as bleeding, spotting, other medical reasons, other non-medical reasons, and so on \citep{said86}. Paradata are also collected even in public survey data. For example,  The National Health Interview Survey(NHIS) releases the data of their survey as Public Use Microdata Files(PUMFs) and also does its paradata. In this data, questionnaire may not be obtained  by several causes such as language problem, absence, refusal, and so on. However, there is no theory to analyze this type of data so far. Therefore, we define a new missing indicator, missing-data mechanisms according to a missing cause, and  study their ignobility in this situation.

\begin{figure}[htbp]
  \begin{center}
  \includegraphics[scale=0.8]{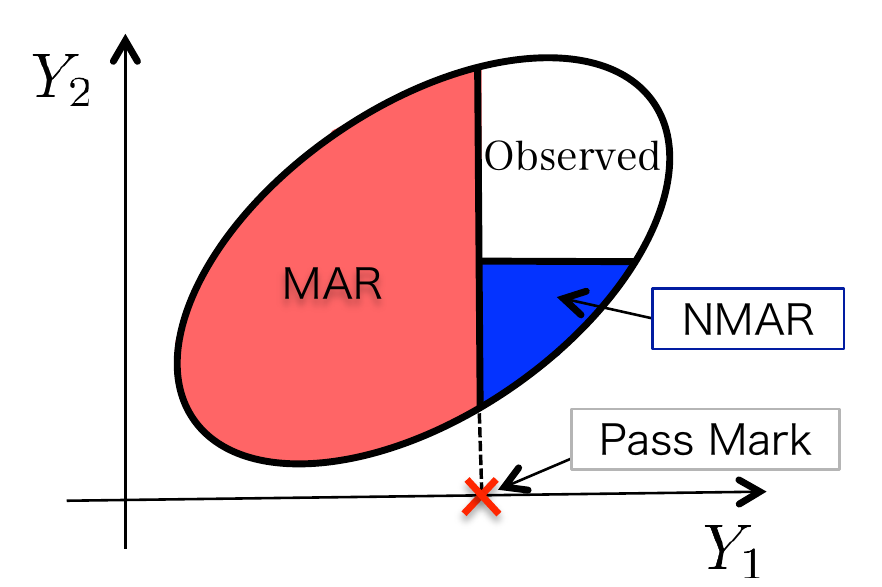}
  \end{center}
  \caption{Illustration of Example: \textit{the relation of the entrance and graduation examination} }
  \label{fig:1}
\end{figure}

For example, consider an example  \textit{Planned versus unplanned missing values} given in \citet{harel09}. Imagine a two-occasion study with a two-stage design in which all subjects are measured at the first wave. Some subjects are chosen for a second wave of measurement by explicit rules that may depend on their recorded responses at the first wave(MAR). However, some of these chosen subjects do not show up at the second wave for causes that may be related to their missing outcomes(NMAR). There are two causes of missingness, but each of it occurs at different waves. In this case, the former cause of missingness is ignorable because of MAR, while the latter cause is non-ignorable because of NMAR. What if these two different causes of missingness  occurs at the same wave? To consider this situation, we give another example; we make a bit more developing the example \textit{the relation of the entrance and graduation examination}. Denote the score of the entrance examination by $Y_1$, that of the graduation examination by $Y_2$. In addition, let the first cause be failed to the entrance examination and the second cause be the absence from the graduation examination since students considered they did not have enough confidence to pass it with their ability at that time. Note that the first cause is MAR, on the other hands, the second cause can be considered as NMAR since, actually, the absent cause  depends on their ability, but it is quite associated with the missing value $Y_2$. This illustration is shown in Figure \ref{fig:1}. Does the first cause of missingness(MAR) become ignorable and the second cause of missingness(NMAR) become non-ignorable as is the usual case when the number of  cause of missingness is only one?  There are no methods to deal with this type of data so far and this paper was intended as an attempt to motivate analyzing it and study its ignorability such as the combination of ``MAR \& MAR'' or  ``MAR \& NMAR''. 

\section{Methods of Handling Missing-data with Different causes}

\subsection{Definition of Partially Missing at Random}
\citet{harel09} argued ways to formalize the notion that only part of the missingness is ignorable that is called partially missing at random(PMAR). PMAR is a condition which gives partial ignobilities when data are missing by different causes, but all subjects may be missing by only one reason at one time like the example of \textit{Planned versus unplanned missing values} and  an assumption to a missing-data mechanism in which not the whole mechanism but the partial mechanism is assumed that it is independent on missing values. \\

\noindent {\bf Definition.} {\it PMAR}\\
\label{def:1}
Let $k$ be a  coarsened summary or many-to-one function of $\bm{M}$ and suppose that a missing-data mechanism is factorized as 
\begin{align*}
P(\bm{M}=\bm{m}\mid \bm{y};\,\bm{\tau})=P(k(\bm{M})=k(\bm{m})\mid \bm{y};\,\bm{\gamma})P(\bm{M}=\bm{m}\mid \bm{y},\,k(\bm{m});\,\bm{\delta}),
\end{align*}
where $\bm{\tau}=[\bm{\gamma}',\,\bm{\delta}']'$ and supposed PD on $\bm{\gamma}$ and $\bm{\delta}$. The missing-data mechanism is said to be PMAR given $k(\bm{M})$ if
\begin{align*}
P(\bm{M}=\bm{m}^{(\ell)}\mid \bm{y},\,k(\bm{m}^{(\ell)});\,\bm{\delta})=P(\bm{M}=\bm{m}^{(\ell)}\mid \bm{y}^{(\ell)},\,k(\bm{m}^{(\ell)});\,\bm{\delta})
\qquad \mathrm{for}\ \forall\bm{\delta},~\forall\bm{y},~\forall \ell.
\end{align*}

If a missing-data mechanism is PMAR, what we have to model is reduced $P(\bm{M}=\bm{m}\mid \bm{y})$ to $P(c(\bm{M})=c(\bm{m})\mid \bm{y})$. Actually, in a similar way of the proof of MAR's ignorability, 
\begin{align*}
L_N(\bm{\xi})&=\prod_{\ell=0}^L\prod_{i\in I_{\ell}}\int  g(\bm{m}^{(\ell)},\,\bm{y}_i\mid\bm{\xi}) d\bm{y}^{(-\ell)}_i\\
&=\prod_{\ell=0}^L\prod_{i\in I_{\ell}}\int  P(k(\bm{M})=k(\bm{m}^{(\ell)})\mid \bm{y}_i;\,\bm{\gamma})P(\bm{M}=\bm{m}^{(\ell)}\mid \bm{y}_i,\,k(\bm{m}^{(\ell)});\,\bm{\delta})f(\bm{y}_i;\,\bm{\theta})d\bm{y}^{(-\ell)}_i\\
&=\prod_{\ell=0}^L\prod_{i\in I_{\ell}}\int  P(k(\bm{M})=k(\bm{m}^{(\ell)})\mid \bm{y}_i;\,\bm{\gamma})P(\bm{M}=\bm{m}^{(\ell)}\mid \bm{y}^{(\ell)}_i,\,k(\bm{m}^{(\ell)});\,\bm{\delta})f(\bm{y}_i;\,\bm{\theta})d\bm{y}^{(-\ell)}_i\\
&=\prod_{\ell=0}^L\prod_{i\in I_{\ell}} P(\bm{M}=\bm{m}^{(\ell)}\mid \bm{y}^{(\ell)}_i,\,k(\bm{m}^{(\ell)});\,\bm{\delta})\int  P(k(\bm{M})=k(\bm{m}^{(\ell)})\mid \bm{y}_i;\,\bm{\gamma})f(\bm{y}_i;\,\bm{\theta})d\bm{y}^{(-\ell)}_i,
\end{align*}
where $\bm{\xi}=[\bm{\theta}',\,\bm{\delta}',\,\bm{\gamma}']'$ and the assumption of PMAR is used at third equation. Therefore, the parameter $\bm{\delta}$ is separable from $\bm{\theta}$, but $\bm{\gamma}$ is not. However, as stated above, this result covers the example of \textit{Planned versus unplanned missing values}, but does not the example of \textit{the relation of the entrance and graduation examination}. This result is ascribable to the limit of missing-indicator $\bm{M}$, which is a binary vector and represents only corresponding $\bm{Y}$ is whether observed or not; $\bm{M}$ does not have information about the cause of missingness. Hence, $\bm{M}$ needs to be extended to have both of them.  In this paper, we consider the situation when data are missing by several different causes at one time, i.e., this is a different situation from that of considered in PMAR. 

\subsection{Extension of  Missing-data Indicator and  Mechanism}
We extend a missing indicator $\bm{M}$ from a binary random variable to a categorical random variable in which one category corresponds to one cause of the missingness, i.e., if there are $C$ causes of missingness, define missing-indicator as
\begin{align*}
  M_t = \begin{cases}
    0 & (Y_t~\mathrm{is}~\mathrm{observed}) \\
    1 & (Y_t~\mathrm{is}~\mathrm{missing}~\mathrm{by~the~cause~}1)\\
    \vdots & \\
    C & (Y_t~\mathrm{is}~\mathrm{missing}~\mathrm{by~the~cause~}C).
  \end{cases}
\end{align*}
When data are missing by the same cause but occurred at different waves such as a side effect, we think it as different causes in this paper. Hence, one missing cause can occur at one wave. By extending a missing indicator like this, different missing-data mechanisms can be defined according to a missing cause.\\

\noindent {\bf Definition.} {\it Missing at random by the cause``$c$"}\\
A missing-data mechanism of a cause ``$c$'' $(c=1,\,\ldots,\,C)$ is called missing at random, if 
\begin{align*}
P(\bm{M}=\bm{m}^{(\ell)}\mid \bm{y};\,\bm{\tau})=P(\bm{M}=\bm{m}^{(\ell)}\mid \bm{y}^{(\ell)};\,\bm{\tau})\qquad \mathrm{for}\ \forall\bm{\tau},~\forall\bm{y},~\ell\in\mathcal{L}_c
\end{align*}
holds, where $\mathcal{L}_c:=\{\ell\in\{0,\;1,\;\,\ldots,\;L\}\mid \mathrm{The~cause~of~missingness~is~}c\,(c=1,\,\ldots,\,C)\}$. Define MCAR and NMAR in the same way.

\begin{example} \textit{The example of entrance and graduation examination}\\
Denote the first cause which is failed to the entrance examination by $M_2=1$ and the second cause which is absent from the graduation one by $M_2=2$. In this case, there are two missing causes so that $C=2$,  $L=2$, $\mathcal{L}_1=\{1\}$ and $\mathcal{L}_2=\{2\}$. Moreover, each missing patter becomes $\bm{m}^{(0)}=(0,\;0)$, $\bm{m}^{(1)}=(0,\;1)$, and $\bm{m}^{(2)}=(0,\;2)$. In the sequel, we use only $M_2$ not $\bm{M}=[M_1,\,M_2]'$ since we assume $M_1=0$ always holds.  The definition leads to the fact that the cause ``1'' becomes MAR and ``2'' becomes NMAR since, as stated at above Definition \ref{def:1}, 
\begin{align*}
P(M_2=1\mid y_1,\,y_2)&=P(M_2=1\mid y_1)=P(M_2=1\mid \bm{y}^{(1)}),\\
P(M_2=2\mid y_1,\,y_2)&=P(M_2=2\mid \bm{y}^{(2)},\,\bm{y}^{(-2)})
\end{align*} 
holds.
\end{example}
The definition of above missing-data mechanisms yields next two propositions. 
\begin{proposition} \label{prop:1}
\textit{The ignorability of ``MAR \& MAR''}\\
If the missing-data mechanisms are MAR for all $c$, all of them are ignorable, i.e., Direct Likelihood can produce consistent estimators.
\end{proposition}
\begin{proposition}\label{prop:2}
\textit{The non-ignorability of ``MAR \& NMAR''}\\
If the missing-data mechanism is NMAR for some $c$, all of them are not ignorable generally.
\end{proposition}
Proposition \ref{prop:1} is similar with the ordinal result when the total number of the ignorable missing causes is one and it is very easy to show. Proposition \ref{prop:2} means if even one NMAR missing-data mechanism exists, all of them must be specified correctly and taken into the model: even MAR missing-data mechanisms can not ignore.  The proof is given in Appendix \ref{sec:a1}. However, there is an exception to make some missing-data mechanisms ignorable, that is  ``{\it hierarchical structure}'' of missing causes.\\

\noindent {\bf Defnition.} {\it Priority and Hierarchical Structure of missing causes}\\
The missing cause ``$c$"  is said to have stronger missing priority than ``$b$" when subjects only who are not missing by the cause ``$b$" can be missing by the cause ``$c$". Here, assume that priority order is given as $``1">``2">\ldots>``C"$ without loss of generality; subjects can be missing by the cause ``$b$" only when not missing by the stronger-priority cause $``c"\;(\forall ``c">``b")$. When observed or missing is decided according to such a priority order of missingness, we describe this system of mechanism as  hierarchical structure. When missing causes have hierarchical structure, we can obtain next propositions.

\begin{proposition} \label{prop:3}
\textit{Hierarchical structure when missing causes are known}\\
When missing-data mechanisms have a hierarchical structure with priority order $``1">``2">\ldots>``C"$ and following condition \eqref{2.1} holds for some $c$, we can ignore the missing cause $c$.  When data can be missing by the cause ``$c$" at $t$ wave, 
\begin{align}
\begin{split}
&P(M_t=c\mid M_{t-1}=0,\;M_t\neq b(\forall ``b">``c"),\; \bm{y};\;\bm{\tau}_c)\\
&=P(M_t=c\mid M_{t-1}=0,\;M_t\neq b(\forall ``b">``c"),\; \bm{y}^{(\ell)};\;\bm{\tau}_c),\qquad \forall\bm{\tau}_c,~\forall\bm{y},
\end{split}
\label{2.1}
\end{align}
where $\bm{\tau}_c$ is a parameter vector which prescribes the missing-data mechanism of the cause $c$.
\end{proposition}
\begin{proposition} \label{prop:4}
\textit{Hierarchical structure when missing causes are unknown}\\
When missing causes are unknown, we can not ignore the missing cause $c$ even if they have hierarchical structure and \eqref{2.1} hold.
\end{proposition}

\begin{figure}[htbp]
  \begin{center}
  \includegraphics[scale=0.7]{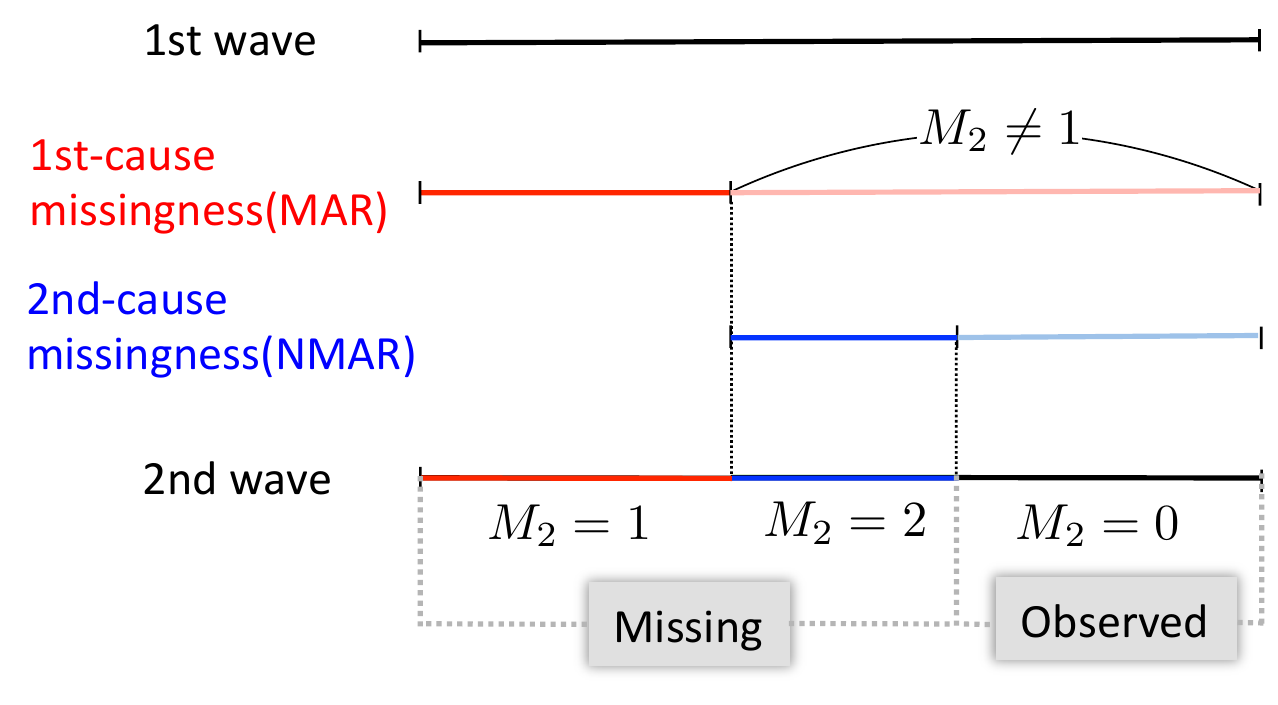}
  \end{center}
  \caption{Illustration of hierarchical structure}
  \label{fig:3.2}
\end{figure}

Proposition \ref{prop:3} mentions that if missing-data mechanisms have hierarchical structure and \eqref{2.1} holds, there exist missing causes to be ignored even if the missing-data mechanisms compose of the combination of non-ignorable missingness (e.g., ``NMAR \& NMAR"). The proof is given in Appendix \ref{a.2}. On the other hand, if missing causes are unknown, this cannot be done, which is proved easily. Thus, when missing causes have hierarchical structure and \eqref{2.1} holds, we can maintain  making some missing-data mechanisms ignorable is a special property of the knowledge of missing causes.
For example, in the example of \textit{The relation of the entrance and graduation examination},  students are divided into two groups at first: $M_2=1$(fail) or $M_2\neq 1$(pass) as is shown in Figure \ref{fig:3.2}. Then, only those who are $M_2\neq 1$ are devided $M_2=0$(present) or $M_2=2$(absent). This is an example of ``hierarchical structure'' of missingness. In this case, the missing-data mechanism of the cause ``1" becomes ignorable missingness by the proposition \ref{prop:3} when missing causes are known, whereas, when they are unknown, we have to specify the correct missing-data mechanism of the cause ``1".

Next, consider the FL in this example. Since, the missing-data mechanism of the cause ``1" can be ignored, but the conditional missing-data mechanism of the cause ``2" $P(M_2=2\mid M_2\neq 1,\;y_1,\;y_2)$ can not be, thus, the probability of observation becomes
\begin{align*}
P(M_2=0\mid y_1,\,y_2;\,\bm{\tau}_1\,\bm{\tau}_2)&=P(M_2\neq 1,\,M_2\neq 2\mid y_1,\,y_2;\,\bm{\tau}_1\,\bm{\tau}_2)\\
&=\{1-P(M_2= 1\mid y_1;\,\bm{\tau}_1)\}\{1-P(M_2= 2\mid M_2\neq 1,\,y_2;\,\bm{\tau}_2)\},
\end{align*} 
where $\bm{\tau}_c\;(c=1,\;2)$ is a parameter vector which prescribes missing-data mechanism of the cause ``$c$". This expression indicates that $\bm{\tau}_1(=\bm{\tau}_{\mathrm{MAR}})$ and $\bm{\tau}_2(=\bm{\tau}_{\mathrm{NMAR}})$ are separable in the observation probability unlike with those in a situation of non-hierarchical structure such as \eqref{a.1}. Then, its FL becomes
\begin{align}
L_N(\bm{\xi})&=\prod_{\ell=0}^{2}\prod_{i\in I_{\ell}}\int  g(\bm{m}^{(\ell)},\,\bm{y}_i\mid\bm{\xi}) d\bm{y}^{(-\ell)}_i\nonumber\\
\begin{split}
&=\prod_{i\in I_{0}}\{1-P(M_2= 1\mid y_{1i};\,\bm{\tau}_1)\}\{1-P(M_2= 2\mid M_2\neq 1;\,y_{2i},\,\bm{\tau}_2)\}f_{Y_1,\,Y_2}(y_{1i},\,y_{2i};\,\bm{\theta})\\
&\quad \times\prod_{i\in I_{1}}P(M_2=1\mid y_{1i};\,\bm{\tau}_{1}) f_{Y_1}(y_{1i};\,\bm{\theta})\\
&\quad\times \prod_{i\in I_{2}} \{1-P(M_2= 1\mid y_{1i};\,\bm{\tau}_1)\}\int  P(M_2=2\mid M_2\neq 1,\,y_{2};\,\bm{\tau}_{2})f_{Y_1,\,Y_2}(y_{1i},\,y_{2};\,\bm{\theta})dy_2.
\end{split}
\label{3.3}
\end{align}
Therefore, we can see that a missing-data mechanism of the cause ``1'' is ignorable and following semi-direct likelihood($SDL_n(\bm{\theta},\,\bm{\tau_2})$)
\begin{align}
\begin{split}
SDL_N(\bm{\xi})&=\prod_{i\in I_{0}}\{1-P(M_2= 2\mid M_2\neq 1,\,y_{2i};\,\bm{\tau}_2)\}f_{Y_1,\,Y_2}(y_{1i},\,y_{2i};\,\bm{\theta}) \\
&\quad \times\prod_{i\in I_{1}}f_{Y_1}(y_{1i};\,\bm{\theta})\\
&\quad \times \prod_{i\in I_{2}} \int  P(M_2=2\mid M_2\neq 1,\,y_{2};\,\bm{\tau}_{2})f_{Y_1,\,Y_2}(y_{1i},\,y_{2};\,\bm{\theta})dy_2
\end{split}
\end{align}
gives same MLE of $\bm{\theta}$ as is given by FL \eqref{3.3}. Hence, the knowledge about the missing-data mechanism of the cause ``1'' is not needed if all missing causes are known. 

\section{Simulation Study}
We study the finite-sample performance of MLE of $\bm{\theta}$ by Monte-Carlo simulation and make comparisons when all causes of missingness are known or not under the previous example. We set three scenarios: (i) the true mechanism of $P(M_2=1\mid y_1)$ is MCAR and correctly specified, (ii) the true mechanism of $P(M_2=1\mid y_1)$ is MCAR and specified as MAR and (iii) the true mechanism of $P(M_2=1\mid y_1)$ is MAR and misspecified as MCAR where suppose the mechanism of $P(M_2=2\mid y_1)$ is correctly specified in all of three scenarios. Let the distribution of $\bm{Y}=[Y_1,\,Y_2]'$ be the normal distribution $N_2(\bm{\mu},\,\Sigma)$ where 
\begin{align*}
\bm{\mu}=[\mu_1,\,\mu_2]',\quad 
\Sigma = 
\begin{bmatrix}
\sigma^2_1 & \rho\sigma_1\sigma_2 \\
\rho\sigma_1\sigma_2 & \sigma^2_2
\end{bmatrix}
\end{align*}
and missing-data mechanisms be
\begin{align}
&(\mathrm{MAR}):\qquad ~~~P(M_2=1\mid y_1;\,\bm{\tau}_1)=\frac{1}{1+\exp\{\tau_{1a}(y_1-\tau_{1b})\}},
\label{3.5}\\ 
&(\mathrm{MCAR}):\qquad P(M_2=1\mid y_1;\,\bm{\tau}_1)=p
\label{3.6}
\end{align}
and
\begin{align*}
P(M_2=2\mid M_2\neq 1,\,y_2,\,;\,\bm{\tau}_2)=\frac{1}{1+\exp\{\tau_{2a}(y_2-\tau_{2b})\}},
\end{align*}
where $\bm{\theta}=[\mu_1,\,\mu_2,\,\sigma_1,\,\sigma_2,\,\rho]'$, $\bm{\tau}_1=[\tau_{1a},\,\tau_{1b}]'$ when missing-data mechanism of the cause 1 is MAR,  $\bm{\tau}_1=p$ when that is MCAR and $\bm{\tau}_2=[\tau_{2a},\,\tau_{2b}]'$.
True values of parameters are given in Table \ref{tb:3.1}.

\begin{table}[htbp]
\caption{True values of parameters}
\label{tb:3.1}
\begin{center}
{\small
\begin{tabular}{cccccccccccc}
\hline
 \multicolumn{5}{c}{$\bm{\theta}^*$}& &\multicolumn{3}{c}{$\bm{\tau}^*_1$} & & \multicolumn{2}{c}{$\bm{\tau}^*_2$} \\ \cline{1-5} \cline{7-9} \cline {11-12}
$\mu^*_1$ & $\mu^*_2$ & $\sigma^*_1$ &$\sigma^*_2$ & $\rho^*$ &  &$\tau^*_{1a}$ & $\tau^*_{1b}$ & $p^*$& & $\tau^*_{2a}$ & $\tau^*_{2b}$\\ \cline{1-5} \cline{7-9} \cline {11-12}
50.0 & 50.0 & 15.0 & 15.0 & 0.60 & & 0.500 & 53.0 & 0.250 & & 0.143(=1/7) & 50.0 \\ \hline
\end{tabular}
}
\end{center}
\end{table}
When all missing causes are known, we can use the semi-direct likelihood $SDL_n(\bm{\theta},\,\bm{\tau}_2)$ to construct MLE of $\bm{\theta}$, which does not need the information about the missing-data mechanism of the cause ``1'' in all scenarios, but when we do not know all of them, we have to use FIML. In the scenario (i), both of methods are correctly specified the missing-data mechanism, thus, we can expect that good estimators can be obtained, but  also expect that estimators when the causes are known becomes a bit more efficient than when the causes are unknown since we have to estimate more parameters ``$p$'' when they are unknown. In the scenario (ii), the true mechanism is \eqref{3.6}, but modeled by using the model \eqref{3.5}. Note that this is not misspecification of true model since if $\tau_{1a}\to 0$ and $\tau_{1a}\tau_{1b}\to \log\{p/(1-p)\}$ holds, 
\begin{align*}
\frac{1}{1+\exp\{\tau_{1a}(y_1-\tau_{1b})\}} \to p.
\end{align*}
Therefore, even if we think the mechanism of the cause one as \eqref{3.5}, but true model is \eqref{3.6}, MLE of $\bm{\theta}$ would tend to $\bm{\theta}^*$. For the stability of estimation, we change the parametrization from model \eqref{3.5} to 
\begin{align}
P(M_2=1\mid y_1;\,\bm{\tau}_1)=\frac{1}{1+\exp(\tau'_{1a}y_1+\tau'_{1b})}
\label{3.7}
\end{align}
in the scenario (ii). In this model, 
\begin{align*}
\tau'_{1a}= 0,\, \tau'_{1b}= -\log\{p/(1-p)\}\, \Leftrightarrow\,\frac{1}{1+\exp(\tau'_{1a}y_1+\tau'_{1b})} = p\quad \mathrm{for}~\forall y_1 .
\end{align*}
Finally, in the scenario (iii), we have completely misspecified the mechanism of the cause ``1'', the estimators when the causes are unknown will be biased. \par
\begin{figure}[htbp]
  \begin{center}
  \includegraphics[scale=0.66]{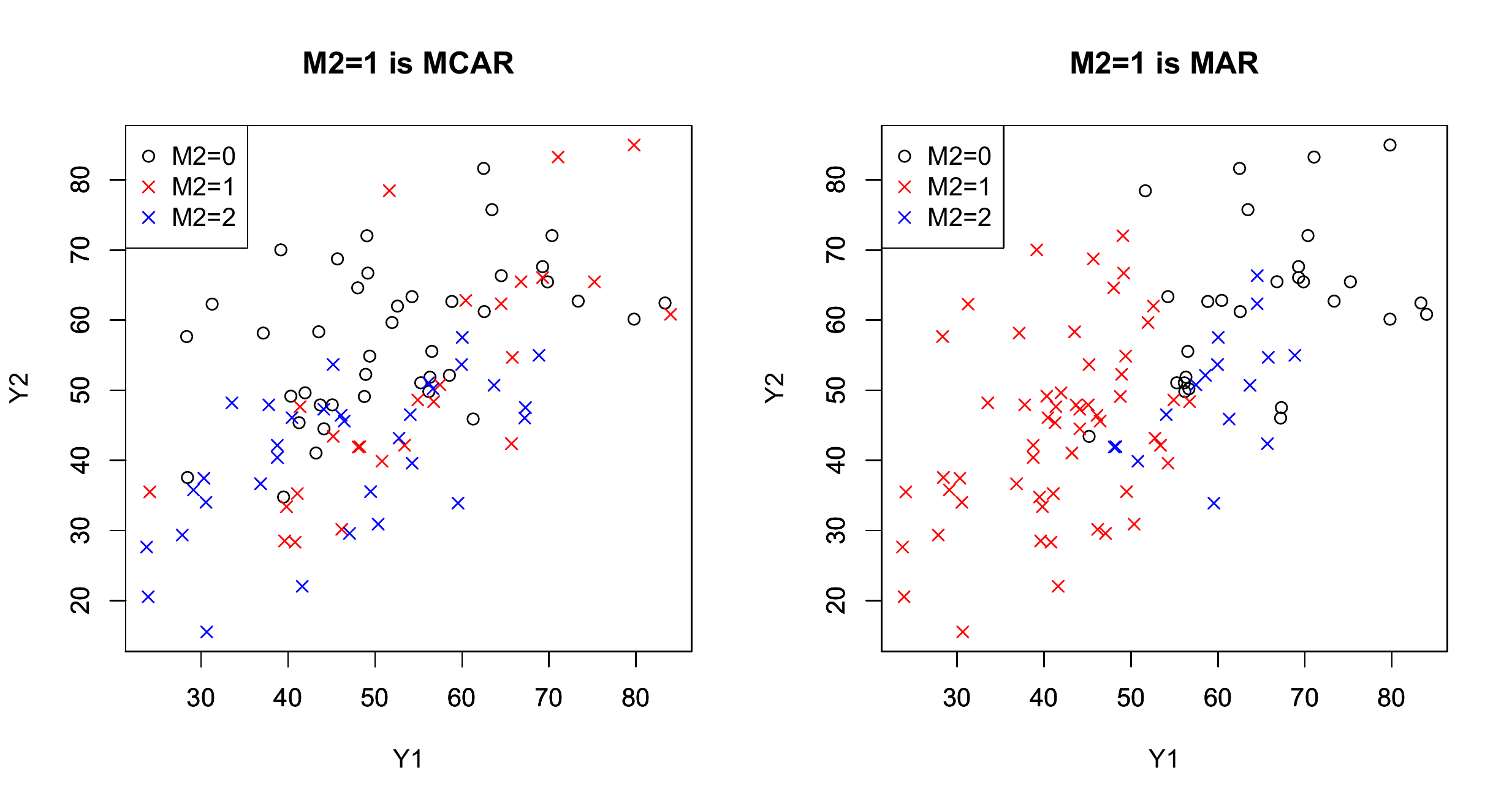}
  \end{center}
  \caption{Plots of a dataset of $[Y_1,\,Y_2]$ with the sample size 100. The left figure is when $P(M_2=1\mid y_1)=0.25$ and the right one is when $P(M_2=1\mid y_1)=\frac{1}{1+\exp\{(y_1-53)/2\}}$.}
  \label{fig:3.3}
\end{figure}

We conducted simulation study with the sample size $n=100$ and plot a dataset in  Figure \ref{fig:3.3}. In this settings, the ratio of $M_2$ approaches ($M_2=0$\,:\,$M_2=1$\,:\,$M_2=2$)=(0.375\,:\,0.250\,:\,0.375) when the mechanism of $M_2=1$ is MCAR and ($M_2=0$\,:\,$M_2=1$\,:\,$M_2=2$)=(0.285\,:\,0.577\,:\,0.138) when that is MAR  as $n$ tends to infinity. R code are given in Appendix \ref{AppendixB} about only scenario (iii).
\subsection{Scenario(i) and Scenario(ii)}
\begin{figure}[htbp]
  \begin{center}
  \includegraphics[scale=0.75]{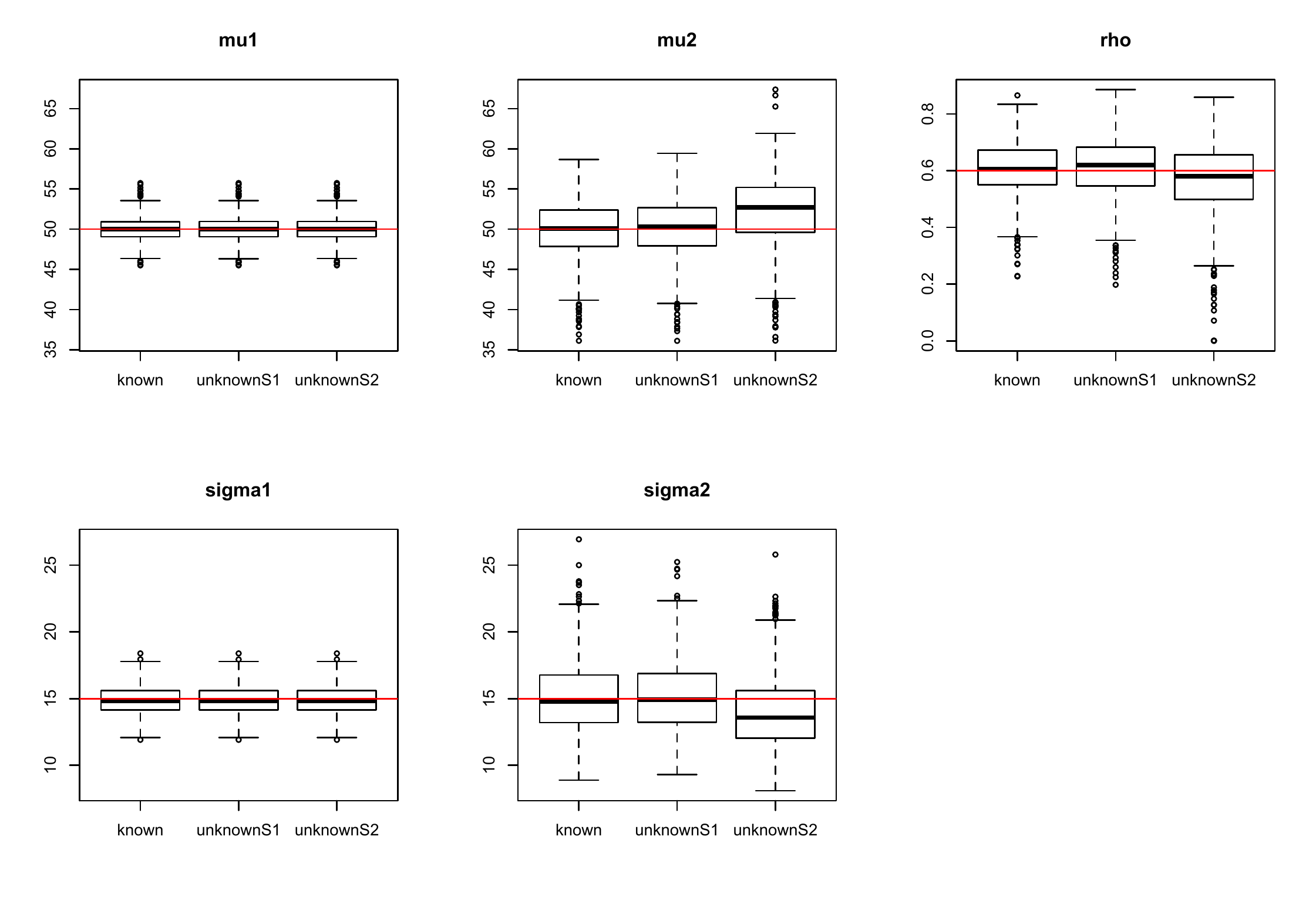}
  \end{center}
  \caption{Box plot in the scenario(i) and scenario(ii). The leftmost ``known'' in each figure means the estimated parameter when the causes of missingness are known, the center ``unknownS1'' means that when they are unknown in the scenario(i) and the rightmost ``unknownS2''  means that when they are unknown in the scenario(ii).}
  \label{fig:3.4}
\end{figure}

Scenario(i) and Scenario(ii) analyze same data sets and we report all together this two scenarios. The box plots of estimated prameters of $\bm{\theta}$ are shown in Figure \ref{fig:3.4} where ("mu1", "sigma1", "rho", "sigma1", "sigma2") is a box plot of $(\mu_1,\,\sigma_1,\,\rho,\,\sigma_1,\,\sigma_2)$ respectively. The estimated parameters of $[\mu_1,\,\sigma_1]$ are all but same in all scenarios since we have complete data on $Y_1$.
In the scenario (i), all parameters are estimated successfully whether we know all causes of missingness or not. Although, one more parameters are estimated when the causes of missingess are unknown than when those are known, there is little difference. In the scenario(ii), the estimated parameters of $[\mu_2,\,\sigma_2]$ may be biased. This is  thought to be causally related to the acute performance of the estimated parameter $\bm{\tau}=[\bm{\tau}_1,\,\bm{\tau}_2']'$. We also calculate root-mean-square error(RMSE) of each parameter and it is shown in Table \ref{tb:3.2} in which ``$\times$" stands for that we did not estimate the concerned parameter since it was not needed. It is seen that the inference on $\tau'_{1a},\,\tau'_{1b}$ does not work well. We thought that weakness of the identification is the primarily cause. This case is close to mixed distribution; It is well-known that the identification of parameters which prescribe a mixed distribution is weak. In fact, the estimated curve of the missing-data mechanism of the cause ``1'' often became a logistic curve similar to that of the cause ``2''.  The cause is conjectured that some subjects who were missing by the cause ``2'' were recognized that they were missed by the cause ``1'' wrongly and then, the missing-data mechanism of the cause ``1'' were also inferred like NMAR. There may be a suspicion of unidentifiability of the model, but  we made sure numerically that this model has identifiability, which is discussed particularly in Chapter 4. Therefore, even if supposed model contains a true model, the inference may go the wrong way.
\subsection{Scenario(iii)}
\begin{figure}[htbp]
  \begin{center}
  \includegraphics[scale=0.75]{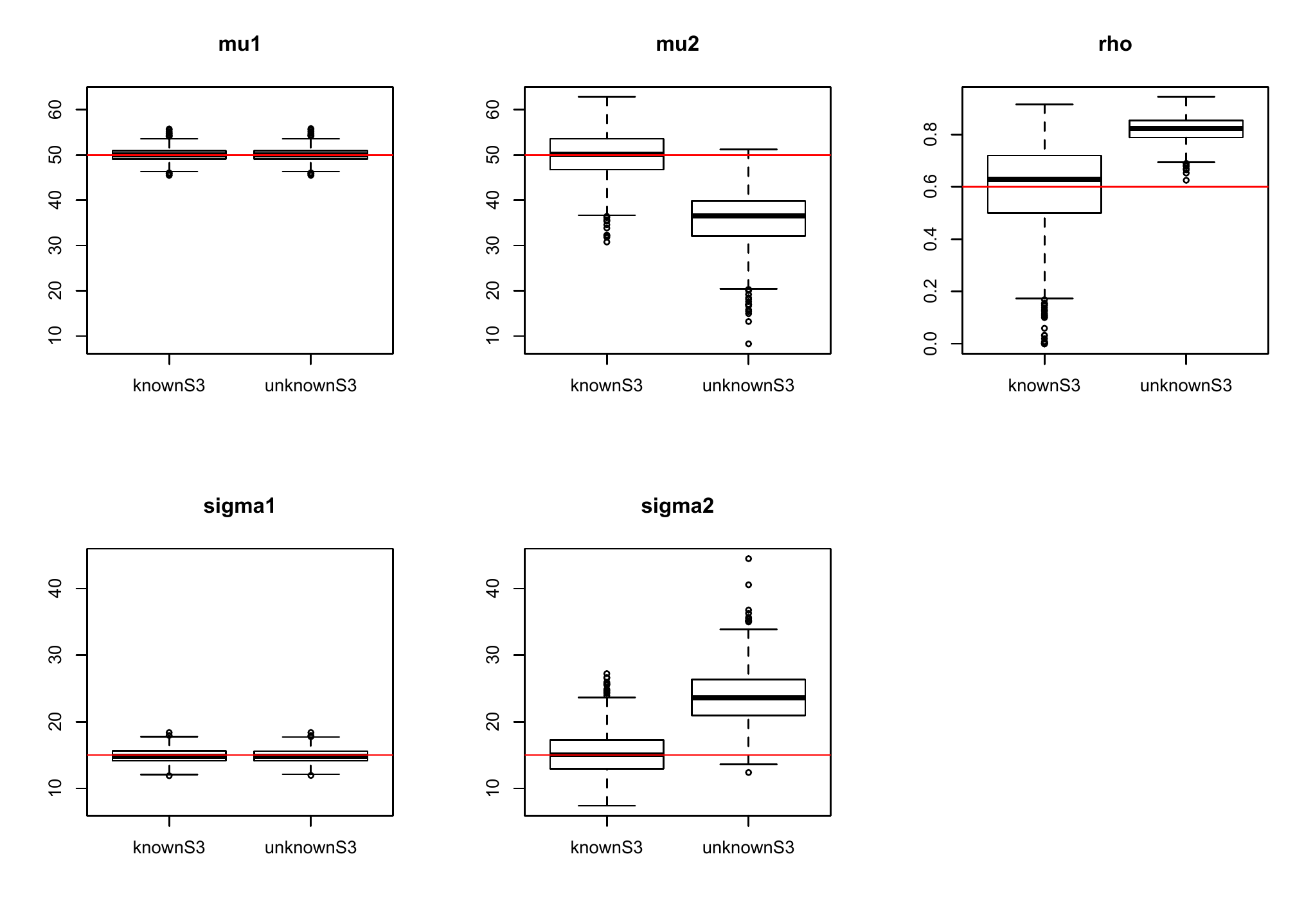}
  \end{center}
  \caption{Box plot in the scenario(iii). The left ``knownS3'' in each figure means the estimated parameter when the causes of missingness are known,  and the right ``unknownS3''  means that when they are unknown in the scenario(iii).}
  \label{fig:3.5}
\end{figure}

Unlike the previous two senarios, supposed model does not contain a true model and we can entertain the following results unsurprisingly. Box plots and RMSE of the result are shown in Figure \ref{fig:3.5} and Table \ref{tb:3.3}. Hence, whether we know the causes of missingess or not give considerable good performance to the estimators. Therefore, the misspecification of the missing-data mechanism of cause ``1'' (MAR) provokes a bad performance of inferences even if we specify the true model of the missing-data mechanism of cause ``2'' (NMAR). 
\begin{table}[htbp]
\caption{RMSE of estimated parameters in the scenario(i) and scenario(ii). }
\label{tb:3.2}
\begin{center}
{\small
\begin{tabular}{ccccccccccccc}
\hline
 &\multicolumn{5}{c}{$\bm{\theta}$}& &\multicolumn{3}{c}{$\bm{\tau}_1$} & & \multicolumn{2}{c}{$\bm{\tau}_2$} \\ \cline{2-6} \cline{8-10} \cline {12-13}
&$\mu_1$ & $\mu_2$ & $\sigma_1$ &$\sigma_2$ & $\rho$ &  &$\tau'_{1a}$ & $\tau'_{1b}$ & $p$& & $\tau_{2a}$ & $\tau_{2b}$\\ \cline{1-6} \cline{8-10} \cline {12-13}
known&1.173 & 2.754 & 0.884 & 2.043 & 0.074 & & $\times$ & $\times$ & $\times$ & & 0.855 & 2.999 \\ \hline
unknownS1&1.174 & 2.915 & 0.884 & 2.084 & 0.081 & & $\times$ & $\times$ & 0.189 & & 12.352 & 6.587 \\ \hline
unknownS2&1.173 & 4.026 & 0.883 & 2.381 & 0.099 & & 8.231& 105.667 & $\times$ & & 8.889 & 11.747 \\ \hline
\end{tabular}
}
\end{center}
\end{table}
\section{Discussion and Summary}
There are many cases where every subject is missing for every cause in practice.  There may be also cases where we could obtain all causes of missingness if tried to do. To describe a missing-data mechanism of such data, we want to get the information about the causes of missingness if we can. However, if we had that knowledge, we could not utilize such information with the ordinal methods when the cause of missingness is only one. Therefore, it is necessary to develop the method for analying this type of missing data. 

We extended a missing indicator from a binary random variable to a categorical random variable in which one category corresponds to one cause of the missingness. This definition of missing indicator allows us to define each missing-data mechanism by each cause. In missing data analysis, it becomes often problem whether this mechanism is ignorable or not. If there exists one cause of missingness at one wave, the mechanism MAR (or MCAR) means ignorable mechanism and NMAR means non-ignorable one. We showed that if there are different causes of missingness and its combination is ignorable combinations ``MAR \& MAR'', all missing-data mechanisms are ignorable. While if the combination is ignorable and non-ignorable combination  ``MAR \& NMAR'', all missing-data mechanisms become non-ignorable in general. Thus, all missing-data mechanisms must be correctly specified if there is even single NMAR cause of missingness. However, if the missing causes have ``hierarchical structure'', which means each cause of missingness has the order of priority and whether observed or missing is decided according to high-priority cause by rotation, MAR mechanisms of missingness become ignorable even if there exists NMAR mechanisms of missingness. 

\begin{figure}[htbp]
  \begin{center}
  \includegraphics[scale=0.70]{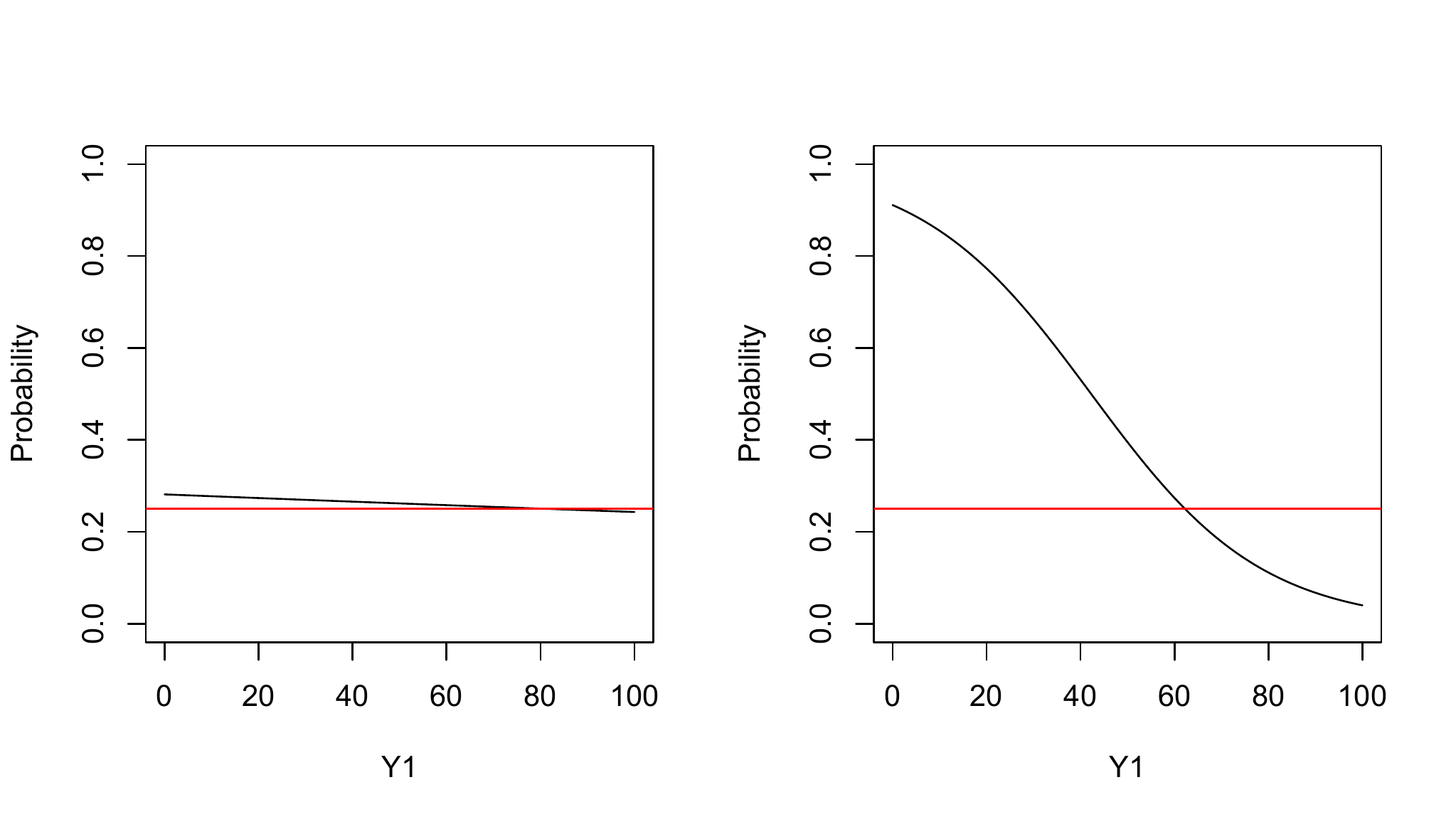}
  \end{center}
  \caption{Two estimated missing-data mechanisms are shown, where true mechanism is $P(M_2=1\mid y_1)=0.25$(MCAR), which is plotted as a red line.}
  \label{fig:4.1}
\end{figure}

Furthemore, we showed the knowledge of all causes of missingness gave considerable good effects to the estimators by simulation study in which we set three scenarios varying true model and supposed model. If we did not know every missing cause, misspecified MAR mechanism severely biased estimators even if NMAR mechanism was correctly specified. Moreover, if supposed MAR mechanism contained true mechanism(MCAR), the identifiability of estimated parameters became weak and the estimators which were not nuisance parameters were also biased. In fact, if we raised the sample size from 100 to 5000, some parameters which regulated MAR missing-data mechanism came close to the true mechanism(MCAR), others still became like NMAR mechanism, but the rate correctly estimated as MCAR mechanism was increased. Two estimated mechanisms of the MCAR mechanism with two different data sets are shown in Figure \ref{fig:4.1}. In the left data set, estimations could be done well, but  in the right one, it could not be. This result shows even if supposed mechanism(MAR) contains a true mechanism(MCAR) and we have enough large sample size, the estimated mechanism may fail to infer the true mechanism because of the weakness of identifiability, while if we know all causes of missingness, such a problem does not occur. We expect taking the information of the missing causes into the model will make the standard deviation of estimators decrease compared to those does not use the information.

\appendix 
\section{Proof of Proposition \ref{prop:2}}
\label{sec:a1}
We prove the proposition \ref{prop:2} for $T=2$; it is proved same way when $T\geq 3$. 
\begin{proof}
Denote a set of missing causes whose missing-mechanism is MAR by $\mathcal{L}_{\mathrm{MAR}}$ and that of NMAR by $\mathcal{L}_{\mathrm{NMAR}}$, i.e., $\mathcal{L}_{\mathrm{MAR}}\cup\mathcal{L}_{\mathrm{NMAR}}=\mathcal{L}=\{1,\,2,\,\ldots,\,C\}$. In addition, let each $\bm{\tau}_{{\mathrm{MAR}}}$ and $\bm{\tau}_{{\mathrm{NMAR}}}$ be the parameter when the missing-data mechanism is MAR and NMAR respectively and thus, $\bm{\tau}=(\bm{\tau}_{{\mathrm{MAR}}},\;\bm{\tau}_{{\mathrm{NMAR}}})$. Suppose parameter distinctness on $\bm{\tau}_{{\mathrm{MAR}}}$ and $\bm{\tau}_{{\mathrm{NMAR}}}$. Then, the probability of observation is defined as
\begin{align*}
P(M_2=0\mid y_1,\,y_2;\,\bm{\tau})&=1-\sum_{\ell=0}^{C}P(M_2=\ell \mid y_1,\,y_2;\,\bm{\tau})\nonumber \\
&=1-\sum_{\ell\in\mathcal{L}_{\mathrm{MAR}}}P(M_2=\ell \mid y_1;\,\bm{\tau}_{\mathrm{MAR}})
-\sum_{\ell\in\mathcal{L}_{\mathrm{NMAR}}}P(M_2=\ell \mid y_1,\,y_2;\,\bm{\tau}_{\mathrm{NMAR}}).
\end{align*}
Therefore, the FL $L_n(\bm{\xi})$ is defined as
\begin{align}
L_N(\bm{\xi})&=\prod_{\ell=0}^{C}\prod_{i\in I_{\ell}}\int  g(\bm{m}^{(\ell)},\,\bm{y}_i\mid \bm{\xi}) d\bm{y}^{(-\ell)}_i\nonumber\\
&=\prod_{\ell=0}^{C}\prod_{i\in I_{\ell}}\int  P(M_2=m^{(\ell)}_2 \mid \bm{y}_i;\,\bm{\tau})f(\bm{y}_i;\,\bm{\theta})d\bm{y}^{(-\ell)}_i\nonumber\\
\begin{split}
&=\prod_{i\in \mathcal{L}_0}\left\{1-\sum_{\ell\in\mathcal{L}_{\mathrm{MAR}}}P(M_2=\ell \mid y_1;\,\bm{\tau}_{\mathrm{MAR}})
-\sum_{\ell\in\mathcal{L}_{\mathrm{NMAR}}}P(M_2=\ell \mid y_1,\,y_2;\,\bm{\tau}_{\mathrm{NMAR}})\right\}\\
&\quad \times f_{Y_1,\,Y_2}(y_{1i},\,y_{2i};\,\bm{\theta})
\end{split}
\label{a.1}\\
&\quad\times \prod_{\ell\in\mathcal{L}_{\mathrm{MAR}}}\prod_{i\in I_{\ell}}P(M_2=m^{(\ell)}_2\mid y_{1i};\,\bm{\tau}_{{\mathrm{MAR}}}) f_{Y_1}(y_{1i};\,\bm{\theta})\nonumber\\
&\quad \times \prod_{\ell\in\mathcal{L}_{\mathrm{NMAR}}}\prod_{i\in I_{\ell}} \int  P(M_2=m^{(\ell)}_2\mid y_{1i},\,y_{2};\,\bm{\tau}_{{\mathrm{NMAR}}})f_{Y_1,\,Y_2}(y_{1i},\,y_{2};\,\bm{\theta})dy_2
\label{2.2}
\end{align}
where each $f_{Y_1,Y_2}$ and $f_{Y_1}$ is a probability density function of $[Y_1,\,Y_2]'$ and $Y_1$. At first grance, $\bm{\tau}_{\mathrm{MAR}}$ has no effects on MLE of $\bm{\theta}$, but it does not. We can see  $\bm{\tau}_{\mathrm{NMAR}}$ and $\bm{\theta}$ are not separable from  \eqref{2.1} as is the case when the total number of causes of missingness is one. We can see to calculate MLE of $\bm{\tau}_{\mathrm{NMAR}}$, on the other hand, the value of $\bm{\tau}_{\mathrm{MAR}}$ is needed from \eqref{2.1}. Therefore, we have to specify the correct model of  $P(M_2=m_2\mid y_{1i};\,\bm{\tau}_{{\mathrm{MAR}}})$ for $\forall\ell\in\mathcal{L}_{\mathrm{MAR}}$ as well as $P(M_2=m_2\mid y_{1i},\,y_{2};\,\bm{\tau}_{{\mathrm{NMAR}}})$  for $\forall\ell\in\mathcal{L}_{\mathrm{NMAR}}$. We have the conclusion.

\section{Proof of Proposition \ref{prop:2}}
\label{a.2}
Though the condition \eqref{2.1} is very similar with that of PMAR, we can not apply the result of PMAR directly because the missing indicator is different from an ordinal one, which is binary random vector. When the data are missing by the cause ``$c$" at $t$ wave, the missing probability of by the cause ``$c$" can be written as
\begin{align*}
&P(M_t=c,\;M_{t-1}=0\mid y_1,\;\ldots,y_{t-1}; \bm{\tau})\\
&=P(M_t=c\mid M_{t-1}=0,\;y_1,\;\ldots,y_{t-1}; \bm{\tau})\prod_{s=2}^{t-1}P(M_s=0\mid M_{s-1}=0,\;y_1,\;\ldots,\;y_s)\\
&=P(M_t=c,\;M_t\neq b(\forall ``b">``c")\mid M_{t-1}=0,\;y_1,\;\ldots,y_{t-1}; \bm{\tau})\\
&\quad\times\prod_{s=2}^{t-1}P(M_s=0\mid M_{s-1}=0,\;y_1,\;\ldots,\;y_s; \bm{\tau})\\
&=P(M_t=c\mid M_t\neq b(\forall ``b">``c"),\; M_{t-1}=0,\;y_1,\;\ldots,y_{t-1}; \bm{\tau}_c)\\
&\quad \times \int P(M_t\neq b(\forall ``b">``c")\mid M_{t-1}=0,\;y_1,\;\ldots,y_{t-1},\;y_t; \bm{\tau}_{-c})f(y_{t}\mid y_1,\;\ldots,\;y_{t-1};\;\bm{\theta})dy_t\\
&\quad\times\prod_{s=2}^{t-1}P(M_s=0\mid M_{s-1}=0,\;y_1,\;\ldots,\;y_s; \bm{\tau}),
\end{align*}
where $\bm{\tau}$ is  decomposed as $\bm{\tau}:=(\bm{\tau}_c,\;\bm{\tau}_{-c})$ and $f$ is a distribution of $Y_t$ given $Y_1,\;\ldots,\;Y_{t-1}$. Therefore, the parameter $\bm{\tau}_c$ is separate from $\bm{\theta}$, but the others may be not.
The problem left here is whether observed probability can be written by the separate parameters unlike in the \eqref{a.1}. If a missing-data mechanism has hierarchical structure,
\begin{align*}
&P(M_t=0\mid y_1,\;\ldots,y_{t}; \bm{\tau})\\
&=P(M_t\neq 1,\;\ldots,\;M_t\neq C\mid y_1,\;\ldots,y_{t}; \bm{\tau})\\
&=P(M_t\neq C\mid M_t\neq 1,\;\ldots,\;M_t\neq C-1,\;y_1,\;\ldots,y_{t}; \bm{\tau}_C)\\
&\quad \times \cdots \times P(M_t\neq 2\mid M_t\neq 1\;y_1,\;\ldots,y_{t}; \bm{\tau}_2)P(M_t\neq 1\mid y_1,\;\ldots,y_{t}; \bm{\tau}_1)
\end{align*}

Hence, we do not need to model the missing-data mechanism of the cause ``$c$".
\end{proof}

\bibliographystyle{apa}
\addcontentsline{toc}{chapter}{\bibname}
\bibliography{refs}  

\end{document}